\documentclass[conference]{IEEEtran}
\IEEEoverridecommandlockouts
\usepackage{cite}
\usepackage{amsmath,amssymb,amsfonts}
\usepackage{algorithmic}
\usepackage{graphicx}
\usepackage{textcomp}
\usepackage{xcolor}
\usepackage{bbm}

\def\BibTeX{{\rm B\kern-.05em{\sc i\kern-.025em b}\kern-.08em
    T\kern-.1667em\lower.7ex\hbox{E}\kern-.125emX}}

\begin{document}
\bibliographystyle{abbrv} 
%
\title{End-to-End Residual CNN with L-GM Loss Speaker Verification System}


\author{
\IEEEauthorblockN{
Xuan Shi, 
Xingjian Du,
Mengyao Zhu$^{\ast}$\thanks{$^{\ast}$ Corresponding author}}

\IEEEauthorblockA{
School of Communication and Information Engineering, Shanghai University\\
Shanghai, China\\}
\IEEEauthorblockA{shixuan1996@i.shu.edu.cn, \{diggerdu, zhumengyao\}@shu.edu.cn }
}

\maketitle

\begin{abstract}
We propose an end-to-end speaker verification system based on the neural network and trained by a loss function with less computational complexity. 
The end-to-end speaker verification system in this paper  consists of a ResNet architecture to extract features from utterance, then produces utterance-level speaker embeddings, and train using the large-margin Gaussian Mixture loss function.
Influenced by the large-margin and likelihood regularization, large-margin Gaussian Mixture loss function benefits the speaker verification performance. 
Experimental results demonstrate that the Residual CNN with large-margin Gaussian Mixture loss outperforms DNN-based i-vector baseline by more than 10\% improvement in accuracy rate.

\end{abstract}


\IEEEpeerreviewmaketitle

\section{Introduction}

Speaker verification (SV) aims to determine whether an utterance comes from the claimed identity or not.
Verification algorithm may require the speaker to utter a specific phrase (text-dependent) or be agnostic to the audio transcript (text-independent). In text-independent SV, no prior constraints are considered for the spoken phrases by the speaker, which makes it challenging compared to text-dependent scenario.

The conventional speaker verification approach entails using i-vectors\cite{DBLP:journals/taslp/DehakKDDO11} and probabilistic linear discriminant analysis (PLDA)\cite{DBLP:conf/icassp/CumaniPL13}. 
As a supervised learning method, i-vector requires sufficient statistics which are computed from a Gaussian Mixture Model-Universal Background Model (GMM-UBM), followed by a PLDA model to produce verification scores\cite{DBLP:journals/taslp/DehakKDDO11}. 
Recently, inspired by using deep neural network in Automatic Speech  Recognition(ASR)\cite{DBLP:journals/spm/X12a}, other research efforts have been conducted on the application of DNN in speaker verification. 
DNN was used to extract abundant statistics and convert them from high dimension to a low-dimension vector, followed by a PLDA or SVM model trained to provide a classification score.

Recently, \cite{LiMJLZLCKZ17} introduced an end-to-end system trained to discriminate between same-speaker and different-speaker utterance pairs. 
First, a deep neural network is used to extract frame-level features from utterances. 
Then, pooling and length normalization layers generate utterance-level speaker embedding, followed by a classifier to give the different predictions to the utterances.
The model to generate embedding is trained with triplet loss \cite{DBLP:conf/cvpr/SchroffKP15}, which minimizes the distance between embedding pairs from the same speaker and maximizes the distance between pairs from different speakers. 
Based on the deep Residual CNN (ResNet) \cite{HeZRS15} and triplet loss \cite{DBLP:conf/cvpr/SchroffKP15}, it outperformed the i-vector speaker verification system.
Nevertheless, triplet loss is still not effective enough because the cosine distance among triplet features are added as additional loss at each time. 
It inevitably results in slow convergence and instability. 
By carefully selecting the image triplets, the problem may be partially alleviated. 
But it significantly increases the computational complexity and the training procedure becomes inconvenient\cite{DBLP:conf/eccv/WenZL016}.

To circumvent the drawbacks of triplet loss, \cite{DBLP:conf/eccv/WenZL016} introduced the center loss by minimizing the Euclidean distance between the features and the corresponding class centroid. 
However, the main drawback of Euclidean distance is that will result in the inconsistency of distance measurements in the feature space. 
The large-margin Gaussian Mixture (L-GM) loss\cite{LGM2018} was proposed to solve the drawbacks of center loss\cite{DBLP:conf/eccv/WenZL016}. 
L-GM loss adopted the Mahalanobis distance to measure the distance between extracted features and the feature centroid of ground truth class, which contributed L-GM loss has a better performance in inter-class dispension and intra-class compactness.

In this paper, we extend the end-to-end speaker embedding systems proposed in\cite{LiMJLZLCKZ17}, but replace the triplet loss with large-margin Gaussian Mixture (L-GM) loss \cite{LGM2018} to get a better performance in the computational complexity and accuracy. 
We use L-GM loss under the assumption that the embeddings of speaker utterances follow a Gaussian mixture distribution approximately. 

Finally, we evaluate our speaker verification system on dataset VoxCeleb\cite{DBLP:conf/interspeech/NagraniCZ17} to investigate the performance. 
In the best case ($\alpha$ = $1$) of ResNet + L-GM loss system performs better than DNN-based i-vector system, which improves the accuracy of verification by more than 10\%.

We review the different methods for speaker verification tasks in Section \uppercase\expandafter{\romannumeral 2} and discuss the performance of them. In Section \uppercase\expandafter{\romannumeral 3}, we propose the structure of the speaker verification system in this paper and elaborate the L-GM loss mathematically. The Experiments compared with the baseline are shown in Section \uppercase\expandafter{\romannumeral 4}.

\section{Related work}

Traditionally, researchers tend to create Gaussian mixture model(GMM)-based speaker verification systems. The most fundamental GMM-based speaker verification methods include the classical maximum a posteriori (MAP) adaptation of universal background model parameters  (GMM-UBM) \cite{DBLP:journals/taslp/FauveMSBM07}  \cite{DBLP:journals/taslp/YinRK07} \cite{DBLP:journals/taslp/HasanH11} and support vector machine (SVM) modeling of GMM super-vectors (GMM-SVM) \cite{DBLP:journals/taslp/CampbellCGRS07}.

I-vector system was proposed in \cite{DBLP:journals/taslp/DehakKDDO11}, which is also a state-of-art and high-effective system to verify speaker's identity. I-vector-based speaker verification models perform classification using cosine similarity between i-vectors or more advanced techniques such as PLDA, heavy-tailed PLDA , and Gauss-PLDA.

Recently, the solution of speaker verification task has increasingly been considered from the perspective of deep learning approaches. There are several models replace the components of traditional SV systems with DNN or other neural network architectures. For example, \cite{DBLP:conf/icassp/HeigoldMBS16} utilized a Deep Neural Network/Hidden Markov Model Automatic Speech Recognition (DNN/ HMM ASR) system to extract content-related posterior probabilities from utterances. \cite{DBLP:conf/icassp/HeigoldMBS16} proposed an transfer learning method based on Bayesian joint probability to help find a better optimal solution of PLDA parameters for the target domain.

A growing number of papers presented end-to-end neural networks for speaker verification. \cite{DBLP:conf/specom/MalykhNK17} utilized the ResNet with spectrograms as an input features in the text-dependent speaker verification task and similar architecture was used in\cite{LiMJLZLCKZ17} for text-independent speaker verification study.

\section{speaker verification system}

\subsection{Overview}

\begin{figure}[!h]
\centering
\hspace{48pt}
\includegraphics[width=0.8\linewidth]{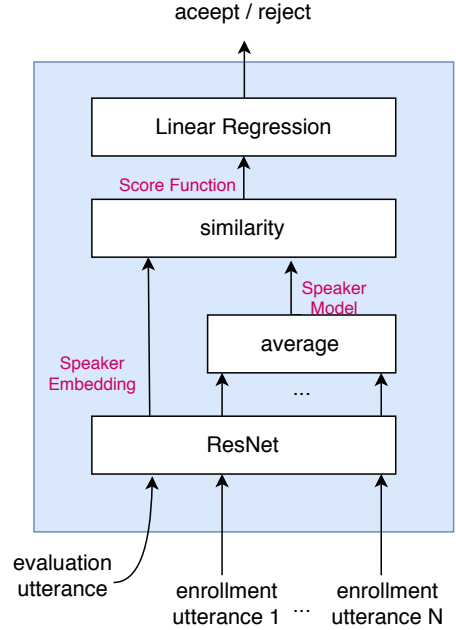}
\vspace{-.5em}
\caption{Full architecture}
\label{fig:pipline}
\vspace{-1em}
\end{figure}

The proposed architecture bases a Residual CNN(ResNet) that extracts features from utterances in training set and maps them to speaker embeddings. 
In this procedure, the objective function deals with embeddings to compact intra-class variations and separable inter-class differences as much as possible.
For enrollment, the speaker model whose parameters has been fixed in training is used to generate embeddings from each utterances. 
Finally, during the evaluation stage, scoring function provides the similarity of utterances between claimed speaker and input. 
Fig.\ref{fig:pipline} shows the overview of our system.

\subsection{Residual CNN}

Deep neural networks perform better than shallow networks in extracting features, but it is not easy to train them. 
Compared with deep neural networks, ResNet\cite{HeZRS15} are easier to optimize and can gain accuracy from considerably increased depth. 
ResNet is comprised by ResBlock, which is defined as:

\begin{equation}
h= F(x,W_{i})+x
\end{equation}

The $x$ and $h$ denote the input and the output of the ResNet block. 
$F$ is the stacked nonlinear layer’s mapping function. 
And $W_{i}$ means the $i$-th weight of the mapping function $F$.
The formulation of $F(x, W_{i}) +x$ can be realized by feedforward neural networks with “shortcut connections”.

\begin{figure}[!h]
\centering
\hspace{48pt}
\includegraphics[width=1.0\linewidth]{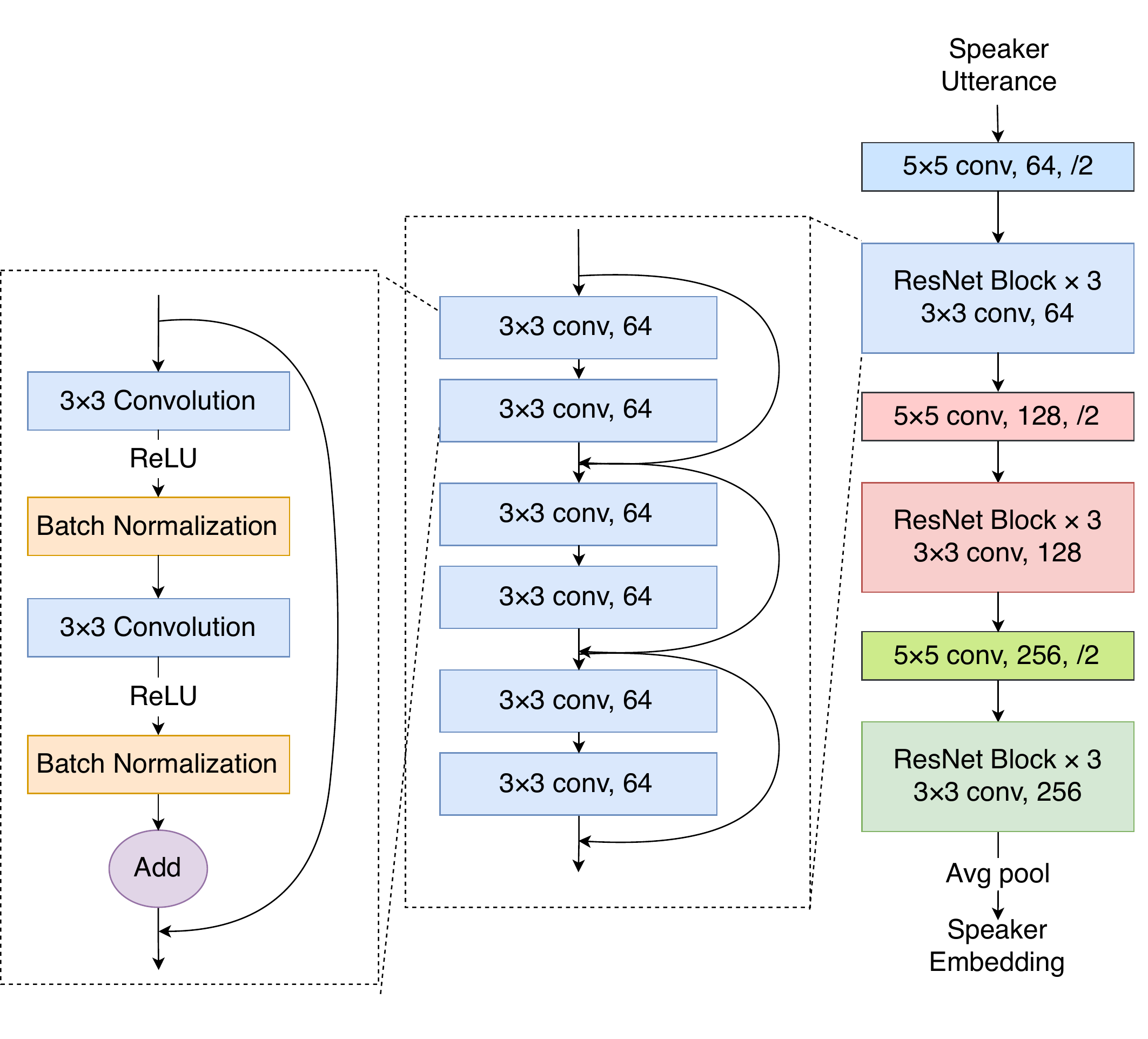}
\vspace{-.5em}
\caption{ResNet architecture}
\label{fig:architecture}
\vspace{-1em}
\end{figure}

Based on the original ResNet architecture, we vary the size of filter and stride for each ResBlock shown in Fig.\ref{fig:architecture}. Thus, each block owns an identical structure and the shortcut is the identity mapping of $x$. Three ResBlocks are stacked in an architecture and the number of channels double. When the number of channels increases, we use a single convolutional layer with filter size of $5 \times 5$ and a stride of $2 \times 2$.

\subsection{Speaker Embedding}

Although an embedding can be extracted from an utterance as long as well in theory, it is constrained by memory practically.
We adopt a expedient method by extracting embeddings from 20 second chunks so that memory could be exploited and features are sufficient for our model to make decisions.
A single embedding is generated from the entire utterance if it is shorter than 20 seconds.
Enrollment embeddings are extracted from one or more utterances, and averaged to create a speaker-level representation.
Enrolling and evaluating utterances are scored by the distance metric used in the objective function.

\subsection{Large-Margin GM Loss}

In the deep speaker system\cite{LiMJLZLCKZ17}, the triplet loss function\cite{DBLP:conf/cvpr/SchroffKP15} was utilized as loss function in speaker verification task, which indicates the similarity between speaker verification task and face verification task.
In the realm of face verification, L-GM loss\cite{LGM2018} is a more efficient and robust loss function, so it is reasonable to investigate the performance of L-GM loss function in speaker verification task.
According to \cite{LGM2018}, L-GM can not only predict the class for a given input following the standard normal distribution, but also perform well in dealing with features which are different from normal inputs.
The condition mentioned above is also fit to the characteristics of voice, as voiceprint based on the physical configuration of a speaker's mouth and throat keeps steady with slight fluctuation in a certain time, which roughly corresponds to the intuition of Gaussian distribution. 
Similar motif can also be discovered in \cite{DBLP:journals/eaai/LefebvreCBZ17}.

We elaborate the large-margin Gaussian Mixture loss function mathematically according to\cite{LGM2018}.
 
Different from the triplet loss\cite{DBLP:conf/cvpr/SchroffKP15}, we hereby assume that the embedding $x$ on the training set follows a Gaussian mixture distribution expressed in Eq.\ref{eq3}. $\mu_{k}$ and $\Sigma_{k}$ are the mean and covariance of speaker $k$ in the embedding space; and $p(k)$ is the prior probability of speaker $k$.

\begin{equation}
\label{eq3}
p(x) = \sum_{k = 1}^{K} \mathcal{N}(x;\mu_{k},\Sigma_{k})p(k)
\end{equation}

Under such an assumption, the conditional probability distribution of an embedding $x_{i}$ given its class label $z_{i} \in [1, K]$ can be expressed in Eq. \ref{rbfprob2}. Consequently, the corresponding posterior probability distribution can be expressed in Eq. \ref{rbfprob3}.

\begin{equation}
\label{rbfprob2}
p(x_i|z_i) = \mathcal{N}(x_i;\mu_{z_i},\Sigma_{z_i})
\end{equation}
\begin{eqnarray}
\begin{aligned}
\label{rbfprob3}
p(z_i|x_i) = \frac{\mathcal{N}(x_i;\mu_{z_i},\Sigma_{z_i})p(z_i)}{\sum_{k=1}^{K}\mathcal{N}(x_i;\mu_k,\Sigma_k) p(k)}
\end{aligned}
\end{eqnarray}

As such, a \emph{classification loss} $\mathcal{L}_{cls}$ can be computed as the cross-entropy between the posterior probability distribution and the one-hot class label as is shown in Eq.~\ref{eq_lossi}, in which the indicator function $\mathbbm{1}()$ equals $1$ if $z_i$ equals $k$; or 0 otherwise.

\begin{eqnarray}
\label{eq_lossi}
\begin{aligned}
\mathcal{L}_{cls} &= -\frac{1}{N}\sum_{i=1}^{N}\sum_{k=1}^{K} \mathbbm{1}(z_i=k) \log {p}(k|x_i) \\
&= -\frac{1}{N}\sum_{i=1}^{N}\log \frac{\mathcal{N}(x_i;\mu_{z_i},\Sigma_{z_i})p(z_i)}{\sum_{k=1}^{K}\mathcal{N}(x_i;\mu_k,\Sigma_k) p(k)}
\end{aligned}
\end{eqnarray}

However, extracted feature $x_i$ may be far away from the corresponding class centroid $\mu_{z_i}$ while still being correctly classified as long as it is relatively closer to $\mu_{z_i}$ than to the feature means of the other classes. To solve this problem, it is necessary to add a \emph{likelihood regularization} term, defined as the sum of negative log likelihood in Eq.\ref{l_reg}, measuring to what extent the training  samples fit the assumed distribution:

\begin{equation}
\label{l_reg}
\mathcal{L}_{lkd} = -\sum_{i=1}^N\log \mathcal{N}(x_i;\mu_{z_i},\Sigma_{z_i})
\end{equation}
Finally the proposed GM loss $\mathcal{L}_{GM}$ is defined in Eq.~\ref{l_gm}, in which $\lambda$ is a non-negative weighting coefficient.
\begin{equation}
\label{l_gm}
\mathcal{L}_{GM} = \mathcal{L}_{cls} + \lambda \mathcal{L}_{lkd}
\end{equation}

By definition, for the training feature space, the classification loss $\mathcal{L}_{cls}$ is mainly related to its discriminative capability while the likelihood regularization $\mathcal{L}_{lkd}$ is related to its probabilistic distribution.
Under the GM distribution assumption, $\mathcal{L}_{cls}$ and $\mathcal{L}_{lkd}$ share all the parameters.

To optimize the generalization of loss function, large classification margin is applied in the training process.
Denote $x_i$'s contribution to the classification loss to be $\mathcal{L}_{cls,i}$, of which an expansion form is in Eq.~\ref{l_cls_i} and Eq.~\ref{l_cls_d}.

\begin{equation}
\label{l_cls_i}
\mathcal{L}_{cls,i} = -\log \frac{p(z_i)|\Sigma_{z_i}|^{-\frac{1}{2}}e^{-d_{z_i}}}{\sum_{k}p(k)|\Sigma_k|^{-\frac{1}{2}}e^{-d_k}}
\end{equation}
\begin{equation}
\label{l_cls_d}
d_k = (x_i - \mu_k)^T\Sigma_{k}^{-1}(x_i - \mu_k) / 2
\end{equation}

The $d_k$ denotes the squared Mahalanobis distance which is obviously non-negative. Then, a classification margin $m \geq 0$ is added to $d_k$ so that $\mathcal{L}_{cls,i}$ gets large margin.

\begin{equation}
\label{l_cls_m}
\mathcal{L}_{cls,i}^{m} = - \log \frac{p(z_i)|\Sigma_{z_i}|^{-\frac{1}{2}}e^{-d_{z_i}-m}}{\sum_{k}p(k)|\Sigma_k|^{-\frac{1}{2}}e^{-d_k-\mathbbm{1}(k=z_i)m}}
\end{equation}

The  $x_i$ is classified to the class $z_i$ if and only if Eq.~\ref{explain_margin} holds, indicating that $x_i$ should be closer to the feature mean of class $z_i$ than to that of the other classes by at least $m$.

\begin{equation}
\label{explain_margin}
e^{-d_{z_i} - m} > e^{-d_k} \Longleftrightarrow d_k - d_{z_i} > m \quad ,\forall k \ne z_i
\end{equation}

It is in a dilemma to fix the value of margin $m$ properly.
On one hand, a large margin could significantly force the features of different classes apart and pull the features of same class to their feature mean of class.
On the other hand, a large margin may cause a difficulty of optimizing for every class when number of classes gets larger.
Therefore, introducing an adaptive scheme for designing the margin is quite necessary.
 An adaptive scheme is proposed by setting the value of $m$ to be proportional to each sample's distance to its corresponding class feature mean, i.e., $m = \alpha d_{z_i}$, in which $\alpha$ is a non-negative parameter controlling the size of the expected margin between two classes on the training set\cite{LGM2018}.

\section{Experiment}

In this sector we present out experiment setup, as well as details related to the ResNet architecture, L-GM loss, and neural network training.

\subsection{Dataset}

In this paper, we run our speaker verification system on dataset VoxCeleb\cite{DBLP:conf/interspeech/NagraniCZ17} to investigate the performance.
VoxCeleb contains over 100,000 utterances for 1,251 celebrities, extracted from videos uploaded to YouTube. The dataset is gender balanced, with 55\% of the speakers male. The speakers span a wide range of different ethnicities, accents, professions and ages.

\subsection{Experimental Setup}

For verification, all Person of Internet(POIs) whose name starts with an `E' are reserved for testing, since this gives a good balance of male and female speakers.
These POIs are not used for training the network, and are only used at test time.
The statistics are given in Table~\ref{table:tt_ver}.

\begin{table}[h!]
\centering
\footnotesize
\begin{tabular}{| l | r | r | r | }
  \hline
  \textbf{Set} & \# POIs & \# Vid. / POI & \# Utterances \\ \hline
  \textbf{Dev} &   1,211   &   18.0   &  140,664 \\ \hline
  \textbf{Test} &   40  &    17.4   &  4,715 \\ \hline

\end{tabular}
\vspace{3pt}
\normalsize
\caption{Development and test set statistics for verification.}
\label{table:tt_ver}
\end{table}

Two key performance metrics: accuracy ratio (ACC) and equal error rate (EER), are used to evaluate system performance for the verification task, both of them are commonly used in identity verification systems\cite{DBLP:conf/interspeech/NagraniCZ17}.


\subsection{Baselines}

\subsubsection{ResNet with triplet loss}

In this paper, we investigate the effect of L-GM loss in the speaker verification task which is mainly different from \cite{LiMJLZLCKZ17}.
To show the superiority of the L-GM loss in the speaker verification task, we adopt the feature extracting model based on the ResNet same with the proposed method and equipped with triplet loss as the objective function.

\subsubsection{DNN-based i-vector}

We also select the DNN based i-vector system, a typical traditional speaker verification method, to compare with the end to end SV system proposed in this paper.

The DNN i-vector model is built based on\cite{DBLP:conf/icassp/LeiSFM14}. 
A seven-layer DNN with 600 input nodes, 1200 nodes in each hidden layer and 3450 output nodes was trained with cross entropy using the alignments from the HMM-GMM. 
The input layer of the DNN is composed of 15 frames (7 frames on each side of the frame for which predictions are made) where each frame corresponds to 40 log Mel-filterbank coefficients.
The DNN is used to provide the posterior probability in the proposed framework for the 3450 senones defined by a decision tree.


\subsection{Experiments}

In this section, we conduct the speaker verification experiments to investigate the performances of large-margin Gaussian Mixture loss and the influence of $\alpha$.
In our experiments, we empirically set $\alpha$ to 1.0, 0.3, 0.1, 0.01, and 0 for our speaker verification tasks.
We also set the regularization $\lambda$ to a small value so that the  $\mathcal{L}_{lkd}$ can truly benefit the training process and improve the training accuracy.

All experiments are carried out using the Pytorch framework.

\subsection{Results}
The Table~\ref{table:results_ver} shows the performance of DNN i-vector system and end-to-end  speaker verification system using the triplet loss and L-GM loss respectively.
According to the Table~\ref{table:results_ver}, with the increasing of margin parameter $\alpha$ in L-GM loss, the inter-class features differences are enlarged as well as the intra-class features variations are reduced, leading to the more effective discrimination.
When $\alpha = 1$, the ACC has the highest value at 90.26\% and lowest EER value at 2.37\%, which performs competitive with the simplified deep speaker system, although a slight gap still exists between them. 
Nevertheless, L-GM loss is more sensible than triplet loss in dealing with speaker verification and audio separation problems, because different speaker owns different voiceprint that could be implicitly clustered from the embedding.

It is also apparent that ResNet using the triplet loss surpasses the performance of DNN i-vector system a lot by 5.74\% higher in ACC and 4.95\% lower in EER, which indicates that end to end system can improve the accuracy of verify speaker identity.




\begin{table}[h!]
\centering
\footnotesize
\begin{tabular}{| l | r | r | }
  \hline
  \textbf{Metrics} & ACC (\%) & EER (\%) \\ \hline
  \textbf{DNN I-vectors}  & 77.83     & 8.80    \\ \hline
  \textbf{ResNet + triplet loss} & 91.43  & 2.17    \\
  \hline 
  \textbf{ResNet + L-GM Loss($\alpha$ = 0)}&70.29 & 10.32   \\
  \textbf{ResNet + L-GM Loss($\alpha$ = 0.01)} &73.60 &9.59  \\ 
  \textbf{ResNet + L-GM Loss($\alpha$ = 0.1)}& 81.23 & 6.84    \\
  \textbf{ResNet + L-GM Loss($\alpha$ = 0.3)}& 85.78& 3.52    \\
  \textbf{ResNet + L-GM Loss($\alpha$ = 1)}&90.26 & 2.37    \\ \hline
\end{tabular}
\vspace{3pt}
\normalsize
\caption{Results for verification on {\tt VoxCeleb} }
\label{table:results_ver}
\end{table}

\section{Conclusion}
The insight of our work does not claim too much effort to understand: a decent and practical speaker verification system relies on an robust  feature extractor to retrieve the intrinsic and distinctive embedding of speaker's voice and a well-designed metric to evaluation the sparsity of embedding space.  In this work, we investigate the efficiency of large-margin Gaussian Mixture loss for model training and the application of ResNet for feature extraction. The combination of  ResNet and large-margin Gaussian Mixture loss function show a promised performance in our dataset and surpass the baseline system with a considered margin in terms of equal error rate.


\section*{Acknowledgment}

This work was supported by the key support Projects of Shanghai Science and Technology Committee (16010500100).



\bibliographystyle{IEEEtran}
\bibliography{SV}
%

\end{document}